\documentclass[aps,superscriptaddress,tightenlines,eqsecnum,nofootinbib]{revtex4}
\usepackage{graphicx,amssymb,amsbsy,bm,amsmath}
\usepackage{hyperref}
\newcommand{\be}{\begin{equation}}
\newcommand{\ee}{\end{equation}}
\newcommand{\bea}{\begin{eqnarray}}
\newcommand{\eea}{\end{eqnarray}}

\newcommand{\vx}{\ensuremath{\vec{x}}}
\newcommand{\vp}{\ensuremath{\vec{p}}}

\newcommand{\vk}{\ensuremath{\vec{k}}}

\begin{document}
\title{Free streaming in mixed  dark matter}
\author{Daniel Boyanovsky}
\email{boyan@pitt.edu} \affiliation{Department of Physics and
Astronomy, University of Pittsburgh, Pittsburgh, Pennsylvania 15260,
USA}
\date{\today}
\begin{abstract}
Free streaming in a \emph{mixture} of collisionless non-relativistic
 dark matter (DM) particles is studied by solving the linearized Vlasov equation implementing methods from the theory of
 multicomponent plasmas. The mixture includes Fermionic, condensed and non-condensed Bosonic
 particles
 decoupling in equilibrium while relativistic, heavy thermal relics that
 decoupled  when non-relativistic (WIMPs), and sterile neutrinos that
 decouple \emph{out of equilibrium} when they are relativistic. The different components interact
 via the self-consistent gravitational potential that they source.  The
 free-streaming length $\lambda_{fs}$ is obtained from the marginal zero  of the
  gravitational polarization function,  which separates short
 wavelength Landau-damped from long wavelength Jeans-unstable
 \emph{collective} modes. At redshift $z$ we find $ \frac{1}{\lambda^2_{fs}(z)} = \frac{1}{(1+z)}\,
\big[ \frac{0.071}{\textrm{kpc}}\big]^2  \sum_{a}     \nu_a \,
g^\frac{2}{3}_{d,a}\,\left( {m_a}/{\mathrm{keV}}\right)^2 I_a  $,
where $0\leq \nu_a \leq 1$ are the \emph{fractions} of the
respective DM components   of mass $m_a$  that decouple when the
effective number of ultrarelativistic degrees of freedom is
$g_{d,a}$, and $I_a$ are dimensionless ratios of integrals of the
distribution functions which only depend on the microphysics at
decoupling and are obtained explicitly in all the cases considered.
If sterile neutrinos produced either resonantly or non-resonantly
that decouple near the QCD scale are the \emph{only} DM component,
we find $\lambda_{fs}(0) \simeq 7\,\mathrm{kpc} \,(\mathrm{keV}/m)$
(non-resonant), $\lambda_{fs}(0) \simeq 1.73\,\mathrm{kpc}\,
(\mathrm{keV}/m)$ (resonant).   If WIMPs with $m_{wimp} \gtrsim
100\,\mathrm{GeV}$ decoupling at $T_d  \gtrsim 10 \,\mathrm{MeV}$
are present in the mixture with $\nu_{wimp} \gg 10^{-12}$,
$\lambda_{fs}(0) \lesssim 6.5 \times 10^{-3}\,\mathrm{pc}$ is
\emph{dominated} by CDM. If a Bose Einstein condensate is a DM
component its free streaming length is consistent with CDM because
of the infrared enhancement of the distribution function.
\end{abstract}

\pacs{98.80.-k,  95.35.+d, 95.30.Cq }

\maketitle

\section{Introduction}
Candidate dark matter (DM) particles are broadly characterized as
cold, hot or warm depending on their velocity dispersions. The
\emph{concordance} $\Lambda\mathrm{CDM}$ standard cosmological model
emerging from CMB, large scale structure observations and
simulations favors the hypothesis that DM is composed of primordial
particles which are cold and collisionless\cite{primack}. Although
this model is very successful in describing the large scale
distribution of galaxies,  recent observations hint at possible
discrepancies summarized as the ``satellite'' and ``cuspy halo''
problems. In the $\Lambda\mathrm{CDM}$ model the CDM power spectrum
favors  small scales, which become non-linear first and collapse in
a hierarchical ``bottom-up'' manner and dense clumps survive the
mergers in the form of ``satellite'' galaxies. Large-scale
simulations within the $\Lambda\mathrm{CDM}$ paradigm  lead  to an
overprediction of  ``satellite'' galaxies \cite{moore2}, which is
almost an order of magnitude larger than the number of satellites
that have been observed in Milky-Way sized
galaxies\cite{kauff,moore,moore2,klyp,will}.

 Furthermore, large-scale  N-body simulations of CDM clustering     predict
 a   density profile monotonically increasing towards the center of the
halos\cite{dubi,frenk,moore2,bullock,cusps}, with asymptotic
behavior $\rho(r)\sim r^{-\gamma}$ where $1\leq \gamma \lesssim
1.5$\cite{moore,frenk,cusps} which describes accurately clusters of
galaxies, but indicates a divergent cusp at the center of the halo.
In contrast with these results, recent observations seem to indicate
central cores in dwarf
galaxies\cite{dalcanton1,van,swat,gilmore,salucci}, sparking the
``cusps vs cores'' controversy.

Warm dark matter candidates (WDM) have been invoked as possible
solutions to these potential discrepancies\cite{mooreWDM,turokWDM},
these particles feature velocity dispersions intermediate between
CDM and HDM and can relieve the satellite and cuspy halo problems.
The clustering properties of collisionless DM in the linear regime
depend on a fundamental length scale: the free-streaming length,
that determines a cutoff in the power spectrum. Length scales larger
than the free-streaming scale undergo gravitational instability, and
shorter scales are damped.  A \emph{simple estimate} of the
free-streaming length $\lambda_{fs}$ is obtained from the familiar
Jeans' length by replacing the speed of sound by the   velocity
dispersion of the particle. An equivalent estimate is obtained by
computing the distance that the particle travels within a dynamical
(Hubble) time\cite{kt} $\lambda_{fs}\sim \langle
\vec{V}^2\rangle^\frac{1}{2}/ H_0$.

However, a   thorough assessment of DM particles and structure
formation requires a more detailed and reliable determination of the
free-streaming length. The necessity for this  has been recently
highlighted by the recent results of ref.\cite{gao} that suggest
that the first stars form in filaments of the order of the free
streaming scale.

Perturbations in a collisionless system of particles with
gravitational interactions is fundamentally different from fluid
perturbations in the presence of gravity. The (perfect) fluid
equations correspond to the limit of vanishing mean free path. In a
gravitating fluid pressure gradients tend to restore hydrostatic
equilibrium with the speed of sound in the medium and short
wavelength fluctuations are simple acoustic waves. For large
wavelengths the propagation of pressure waves cannot halt
gravitational collapse on a dynamical time scale. The dividing line
is the Jeans length: perturbations with  wavelengths shorter than
this oscillate as sound waves, while perturbations with longer
wavelength   undergo gravitational collapse.

 In a gas of collisionless particles with gravitational interaction
 the situation is different since the mean free path is much larger than the size of the
 system (Hubble radius) and the fluid description is not valid.
 Instead the Boltzmann-Vlasov  equation for the distribution function
 must be solved to extract the dynamics  of  perturbations\cite{peebles,bert}. Just as
 in the case of plasma physics, the linearized Boltzmann-Vlasov
 equation describes \emph{collective} excitations\cite{plasmas}. In
 the case of a collisionless gas with gravitational interactions
 these collective excitations describe particles free-streaming in
 and out of the gravitational potential wells of which they are the
 source. The damping of short wavelength \emph{collective}
 excitations is akin to \emph{Landau damping in
 plasmas}\cite{plasmas} and is a result of the phenomenon of
 \emph{dephasing} via phase mixing in which the particles are out of phase with the
 potential wells that they produce\cite{bt}. This situation is similar to
 Landau damping in plasmas where dephasing between the charged
 particles and the self-consistent electric field that they produce
 lead to the collisionless damping of the collective
 modes\cite{plasmas}. For a thorough review of collective
 excitations  and their Landau damping in gravitational systems see ref.\cite{trema1}.

 Gilbert\cite{gilbert} studied the linearized Boltzmann-Vlasov
 equation in a matter dominated cosmology for non-relativistic
 particles described by an (unperturbed) Maxwell-Boltzmann
 distribution function. In this reference the linearized
 this equation was cast as a Volterra integral equation of the
 second kind which
 was solved numerically. The result of the integration reveals a
 limiting value of the wavevector below which perturbations are
 Landau  damped and above which perturbations grow via gravitational
 instability. Eventually the redshift in the expanding cosmology
 makes wavevectors that are initially damped  to enter the band of
 unstable modes and grow\cite{gilbert}. However, the dividing wavevector  between damped
 and growing modes emerges at very early times during which the
 expansion can be neglected\cite{gilbert}. The results of the
 numerical study were consistent with replacing the speed of sound
 by the Maxwellian velocity dispersion in the Jeans length (up to a
 normalization factor of $\mathcal{O}(1)$). Gilbert's equations were
 used also by Bond and Szalay\cite{bondszalay} in their pioneering
 study of collisionless damping of density fluctuations in an
 expanding cosmology. These authors focused primarily on massive
 neutrinos and solved numerically Gilbert's integral equation but approximated
 the Fermi-Dirac distribution function and provided a fitting
 function for the numerical results.  Gilbert's
 equations were also solved numerically to study dissipationless clustering of
 neutrinos in ref.\cite{bran} but with a truncation of their Fermi-Dirac distribution
 function   and an  analytic fit to the numerical solution of the integral equation was
 provided. Bertschinger and Watts\cite{bertwatts,bert} also studied
 numerically  Gilbert's equation within the context of cosmological perturbations
 from cosmic strings and massive light neutrinos, and more recently similar integral  equations were
   solved  approximately for thermal neutrino relics in ref.\cite{ringwald}.

 Most of the studies of the solutions of the linearized
 Boltzmann-Vlasov equation for collisionless particles addressed one
 single species of DM candidates\footnote{The exception being ref.\cite{bertwatts} wherein
 massive neutrinos decoupled with a Fermi-Dirac distribution  and a cosmic string source $\propto \delta^3(\vec{r})$ were studied, neglecting any radiation component.} and generally in terms
 equilibrium distribution functions or approximations thereof, for
 example Maxwell-Boltzmann for relics that decouple
 non-relativistically or Fermi-Dirac (without chemical potential)
 (or truncations of this distribution) for neutrinos.

 However, it is   likely that DM may be composed of
 \emph{several} species, this possibility is suggested in ref.\cite{palazzo} and
    most extensions of the standard model generally allow several possible candidates, from
    massive weakly interacting   particles (WIMPs) to ``sterile'' neutrinos  ( $SU(2)$ singlets)\cite{kusenko}.

 Furthermore, several possible WDM candidates may decouple \emph{out
 of local thermodynamic equilibrium} (LTE) with distribution functions which may be very different
 from the usual ones in LTE. This is the case for sterile neutrinos
 produced non-resonantly via the Dodelson-Widrow (DW)
 mechanism\cite{dw} or through a lepton-driven MSW
 (Mikheyev-Smirnov-Wolfenstein) resonance\cite{este}.

 \medskip

 {\bf{A. Motivation and goals:}} In this article we study free streaming
 of decoupled collisionless non-relativistic (DM) candidates focusing on \emph{two} aspects:
 \begin{itemize}
\item{ A \emph{mixture} of (DM) particles including CDM and WDM candidates: typical studies of
structure formation invoke either CDM \emph{or} WDM, but it is
likely that \emph{both} candidates are present with different
fractions $\nu$ of the total DM component. In fact most particle
physics extensions beyond the standard model have plenty of room for
  a variety of CDM, WDM or HDM candidates. Thus we allow all of these components, each one contributing
  an arbitrary fraction $\nu$ to the total (DM) content of the Universe. Although the DM candidates are
  collisionless and do not interact directly with each other, they
  interact \emph{indirectly} via the gravitational potential that
  they source. As a result the free-streaming length of the mixture is a non-trivial
  function of the individual free-streaming lengths.   }

\item{Free streaming has mostly been studied in the above references with particles that
decoupled either when ultrarelativistic (as is the case for
neutrinos) or non-relativic as in the case of weakly interacting
massive particles (WIMPs) but generally in local thermodynamic
equilibrium (LTE), namely with Fermi-Dirac, Bose-Einstein or
Maxwell-Boltzmann distributions respectively. We seek to obtain the
corresponding free streaming lengths for particles that decoupled
\emph{in or out} of LTE with \emph{arbitrary} isotropic distribution
functions, without any truncation. This aspect is important for
sterile neutrinos either produced non-resonantly\cite{dw} or
resonantly\cite{este,kusenko} because these particles    decoupled
while relativistic but  \emph{out} of LTE. Therefore, we consider
the most general \emph{mixture} of Fermionic and Bosonic thermal
relics that decouple   when relativistic, including the possibility
of a Bose-Einstein condensate (BEC)\cite{coldmatter},   heavy
non-relativistic thermal relics, as the case of WIMPs, and sterile
neutrinos that decoupled \emph{out of LTE} when ultrarelativistic. }
 \end{itemize}

 In order to carry out this program analytically we first neglect the
cosmological expansion and solve the linearized Boltzmann-Vlasov
equation in the non-expanding case \emph{exactly} by implementing
methods from the theory of multicomponent plasmas\cite{plasmas}. The
neglect of the cosmological expansion is warranted by the detailed
numerical study in refs.\cite{gilbert,bondszalay} wherein it was
found that the dividing wavevector between the Landau damped modes
and the modes that grow under gravitational instability is
insensitive to the expansion, just as in the case of the Jeans
instability where the Jeans wavevector can be extracted in the
non-expanding case include the redshift dependence of the density,
speed of sound and scale factors \emph{a posteriori}\cite{kt}. The
analytic \emph{exact} solution for the free-streaming wave-vector
\emph{today} $k_{fs}( 0)=2\pi/\lambda_{fs}( 0)$ in terms of the full
distribution functions without truncation, \emph{in or out of LTE}
is a main result of this program, one which yields a reliable
determination of free-streaming lengths for mixtures of (DM)
components that decoupled with arbitrary distribution functions.

\medskip

This program is carried out by implementing methods from the theory
of multicomponent plasmas\cite{plasmas}, in particular we obtain the
``gravitational polarization'' function\cite{trema1,bt} for a
\emph{mixture} of (DM) components akin to the dielectric response
function of multicomponent plasmas\cite{plasmas}. The collective
excitations are described by the zeroes of this function in the
complex frequency plane and the free-streaming wave-vector $k_{fs}$
is identified as that wavevector that separates between the
Landau-damped short wavelength modes and the gravitationally Jeans-
unstable long-wavelength modes.

Based on the Liouville evolution of the decoupled distribution
functions and assuming that the expansion of the universe is slow
enough so that it can be treated adiabatically, we provide a scaling
argument   that determines  the following dependence of the free
streaming length on the redshift,  \be \lambda_{fs}(z) =
 {\lambda_{fs}(0)}{\sqrt{1+z}}\,.\ee

\medskip

{\bf{B. Summary of  Results:} } Our main result for the comoving
free streaming length  $\lambda_{fs}(z)$  at redshift $z$  of mixed
(DM) is

 \bea \frac{1}{\lambda^2_{fs}(z)} = \frac{1}{(1+z)}\,
\Big[\frac{0.071}{\textrm{kpc}}\Big]^2 \sum_{\textrm{
species}}&&\Bigg\{ \nu_F \,
g^\frac{2}{3}_{d,F}\,\left(\frac{m_F}{\mathrm{keV}}\right)^2 I_F[u]
+ \nu_s \,
g^\frac{2}{3}_{d,s}\,\left(\frac{m_s}{\mathrm{keV}}\right)^2\, 6.814
+\nu_B \,
g^\frac{2}{3}_{d,B}\,\left(\frac{m_B}{\mathrm{keV}}\right)^2
I_B[x_d,u_d]+ \nonumber \\&& 10^{12}\, \nu_{wimp}\, \,
g^\frac{2}{3}_{d,wimp}\,
\Big(\frac{m_{wimp}}{100\,\textrm{GeV}}\Big)\,\Big(\frac{T_{d}}{10
\,\textrm{MeV}}\Big) \Bigg\}\,,\label{lfsintro}\eea  where $\nu_a$
is the \emph{fraction} of (DM) of each component with $\sum_a
\nu_a=1$, $g_{d,a}$ is the effective number of ultrarelativistic
degrees of freedom at decoupling for each species $(a)$ of mass
$m_a$,  and the functions $I_F,I_B$ are dimensionless ratios of
integrals of the distribution functions of the decoupled particles
which are determined by the microphysics at decoupling. Their
explicit expressions in the cases considered are given in section
(\ref{sec:deco}). The label $F$ refer to \emph{all} possible
Fermions with chemical potential $\mu$ decoupled in LTE  at a
temperature $T_d$ while ultrarelativistic, and  sterile neutrinos
produced non-resonantly via the (DW) mechanism\cite{dw} for which
the chemical potential vanishes and
$I_F[0]=2\ln(2)/3\zeta(3)=0.3844$, and the label $s$ refers solely
to sterile neutrinos produced via a lepton-driven (MSW) resonance
via the mechanism described in ref.\cite{este}. The label $B$
corresponds to condensed or non-condensed Bosons of mass $m$ and
chemical potential $\mu$ that decoupled at temperature $T_d$ while
ultrarelativistic. The function $I_B$ features an infrared
divergence in the limits $\mu/T_d ; m/T_d \rightarrow 0$ or $\mu =
m$ for any value of the mass. This latter case corresponds to the
case of a Bose-Einstein Condensate\cite{coldmatter}. Thus Bosonic
particles that decoupled while ultrarelativistic with $\mu/T_d \ll
1$ \emph{or} that formed a BEC lead to small free-streaming lengths
and behave as CDM. Finally, (WIMPs) are considered to be decoupled
while non-relativistic with a Maxwell-Boltzmann distribution
function.

Eq.(\ref{lfsintro}) clearly shows that (WIMPs) with $m\sim
100\,\mathrm{GeV}$ that decoupled kinetically in LTE  at $T\sim
10\,\mathrm{MeV}$\cite{dominik} dominate all other contributions to
$k_{fs}$ resulting in an extremely small free streaming length
$\lambda_{fs} \sim 6.5 \times 10^{-3}\,\mathrm{pc}$ \emph{unless}
their fractional abundance $\nu_{wimp} \lesssim 10^{-12}$.

 If (DM)
is dominated by sterile neutrinos (produced either resonantly or
non-resonantly) that decoupled near the QCD scale\cite{dw,este} with
$m \sim \mathrm{keV}$, we find that the typical free-streaming
lengths \emph{today} are $\lambda_{fs} \sim 2-7 \, \mathrm{kpc}$,
where the larger value corresponds to the non-resonant and the lower
value to the resonant production mechanisms respectively. The lower
values are  consistent with the recent numerical study of the
formation of the first stars in filamentary structures\cite{gao} and
the values for the ``cores'' extracted from the data in
ref.\cite{gilmore} for dwarf spheroidal galaxies (dSphs) ($r_c \sim
0.5 \,\mathrm{kpc}$). The upper values agree with the cores
extracted from the data in ref.\cite{salucci} for spiral galaxies
($r_c \sim 10 \,\textrm{kpc}$).

\section{Free streaming length  for multicomponent Dark Matter}
We  study DM particles that   decoupled in or out of equilibrium
while relativistic or non-relativistic, but that are
non-relativistic \emph{today}. As argued above, Gilbert's detailed
numerical study\cite{gilbert} confirmed by Bond and
Szalay\cite{bondszalay} shows that the value of $k_{fs}$ can be
extracted from the marginal case between modes that grow under
gravitational instability and those that are Landau damped. For this
marginal case linear perturbations are stationary, and just as in
the case of the Jeans length this marginal value can be reliably
extracted in a non-expanding cosmology, including \emph{a
posteriori}   the scale factor dependence of the various quantities
in the Jeans wavelength which separates the gravitationally stable
and unstable modes\cite{kt}. In section (\ref{subsec:redshift}) we
present arguments that determine the redshift dependence of the free
streaming length under a suitable approximation.

Consider several species of   DM candidates that are
non-relativistic \emph{today} with masses $m_a$ and
  distribution functions $f_a$ where the label $ a=1,2 \cdots$ refers to the different
  components. Each
component $(a)$ obeys the collisionless Boltzmann-Vlasov
equation\cite{peebles,bert,bt},

\be \frac{\partial f_a(\vx,\vp\,;t)}{\partial t}+ \frac{\vp}{m_a}
\cdot \vec{\nabla}_{\vx}f_a(\vx,\vp\,;t)-m_a \vec{\nabla}_{\vx}
\Phi(\vx;t) \vec{\nabla}_{\vp} f_a(\vx,\vp\,;t) =0 \label{vlasov}\ee
where $\Phi$ is the total Newtonian potential which is the solution
of the Poisson equation \be \nabla^2 \Phi(\vx;t) = 4\pi G \sum_a
m_a\,g_a \int \frac{d^3p}{(2\pi)^3} f_a(\vx,\vp\,;t)\;,
\label{Poisson}\ee where $g_a$ is the number of internal degrees of
freedom.

We note that whereas each individual species obeys its own
collisionless Boltzmann-Vlasov equation, the Newtonian gravitational
potential is determined by \emph{all the components} as indicated by
the Poisson equation (\ref{Poisson}). Therefore, although the
different DM components do not interact \emph{directly} they
interact \emph{indirectly} via the self-consistent gravitational
potential since \emph{all} of the DM components act as source of
this potential which enters in   Boltzmann-Vlasov equation of each
component.

Linearizing the Boltzmann-Vlasov equation and writing  \be
f_a(\vx,\vp\,;t) = f_a^{(0)}(p)+f_a^{(1)}(\vx,\vp\,;t) ~~;~~
\Phi(\vx;t)=\Phi^0(\vx;t) + \Phi^{(1)}(\vx;t)\label{perts}\ee where
$f_a^{(0)}(p)$ are the distribution functions of the
  species that \emph{decoupled in or out of LTE}, the \emph{only assumption} is that these are isotropic,
namely only depend on $p=|\vp\,|$. In the non-expanding case the
zeroth order equation requires to invoke the usual ``Jeans swindle''
(see the textbooks\cite{bt}) whereas in the expanding case the
zeroth order equation is solved in terms of the inhomogeneous
gravitational potential which yields  the expanding background
(see\cite{peebles}).

It is convenient to perform a spatial Fourier transform of the
perturbations in a volume $V$ \be f_a^{(1)}(\vx,\vp\,;t) =
\frac{1}{\sqrt{V}} \sum_{\vk} F_a^{(1)}(\vk,\vp\,;t)\,
e^{i\vk\cdot\vx} ~~;~~ \Phi^{(1)}(\vx;t) = \frac{1}{\sqrt{V}}
\sum_{\vk} e^{i\vk\cdot\vx} \varphi^{(1)} (\vk;t) \label{FTperts}\ee
in terms of which the Vlasov and Poisson equations  for the
perturbations become \be \frac{dF_a^{(1)} (\vk,\vp\,;t)}{dt} +
i\frac{\vk\cdot\vp}{m_a}F_a^{(1)}(\vk,\vp\,;t)-i m_a
\vk\cdot\widehat{\vp}~\varphi_1(\vk;t) \frac{df_a^{(0)}(p)}{dp} =0
\,, \label{Vlasopert}\ee \be  \varphi_1(\vk;t) = - \frac{4\pi
G}{k^2} \sum_b m_b\,g_b \int \frac{d^3p}{(2\pi)^3}
F_b^{(1)}(\vk,\vp\,;t) \,. \label{Poispert}\ee Inserting eqn.
(\ref{Poispert}) into eqn. (\ref{Vlasopert}) makes explicit  that
all the DM components are actually \emph{interacting} through the
self-consistent gravitational perturbations which act as a dynamical
self-consistent mean field, as discussed above.

\medskip

 Decaying or growing perturbations
must be treated as an initial value problem, and following the
treatment of Landau damping in plasmas\cite{plasmas} we introduce
the Laplace transform of the perturbations \be
\widetilde{F}_a^{(1)}(\vk,\vp\,;s) = \int_0^\infty e^{-st}
{F}_a^{(1)}(\vk,\vp\,;t)\;dt~~;~~\widetilde{\varphi}^{(1)}(\vk\,;s)
= \int_0^\infty e^{-st} {\varphi}^{(1)}(\vk\,;t)\;dt \,.
\label{lapla}\ee The Laplace transform of the Bolztmann-Vlasov
  equation  (\ref{Vlasopert}) leads to  \be
\widetilde{F}_a^{(1)}(\vk,\vp\,;s) =
\widetilde{\varphi}^{(1)}(\vk\,;s) \, \frac{i m_a
\vk\cdot\widehat{\vp} }{s+ i\frac{\vk\cdot\vp}{m_a}}~
\frac{df_a^{(0)}(p)}{dp} + \frac{F_a^{(1)}(\vk,\vp\,;t=0)}{s+
i\frac{\vk\cdot\vp}{m_a}}\,. \label{Fone}\ee Taking the Laplace
transform of (\ref{Poispert}), multiplying  (\ref{Fone}) by $(-4\pi
G m_a\,g_a/k^2)$, summing over $a$ and integrating in $p$ we are led
to \be \widetilde{\varphi}^{(1)}(\vk\,;s) = \frac{i\,4\pi
G}{k^2\,\varepsilon(k;s)} \, \sum_a m_a \,g_a\int
\frac{d^3p}{(2\pi)^3}\frac{F_a^{(1)}(\vk,\vp\,;t=0)}{
 \frac{\vk\cdot\vp}{m_a}-is} \label{varfi2}\ee where the gravitational ``polarization'' function is given by \be
 \varepsilon(k;s) = 1+ \frac{4\pi G}{k^2}\sum_a  m^2_a\,g_a \int
\frac{d^3p}{(2\pi)^3} \frac{ \vk\cdot \hat{\vp}}{
 \frac{\vk\cdot\vp}{m_a}-is}\,\frac{df_a^{(0)}(p)}{dp}
 \label{varepsilon}\ee The collective excitations of the
 collisionless self-gravitating system correspond to the poles of
 $\widetilde{\varphi}^{(1)}(\vk\,;s)$ in the complex s-plane, these
 are the zeroes of  $\varepsilon(k;s)$. The time dependence of the
 perturbation of the Newtonian gravitational potential is obtained by the inverse
 Laplace transform \be \varphi^{(1)}(\vk;t) = \int_C \frac{ds}{2\pi i}\,
 e^{st}\,\widetilde{\varphi}^{(1)}(\vk\,;s)\ee where $C$ stands for
 the Bromwich contour parallel to the imaginary axis and to the
 right of all the singularities of
 $\widetilde{\varphi}^{(1)}(\vk\,;s)$ in the complex $s$-plane.

 Using that $f_a^{(0)}(p)$ is a function of $p=|\vp\,|$ it is
 convenient to write \be \frac{\vk\cdot \hat{\vp}}{\frac{\vk\cdot\vp}{m_a}-is}  =\frac{m_a}{p}\left[1
+\frac{is}{\frac{\vk\cdot\vp}{m_a}-is}\right]  \ee which allows to
extract the $s=0$ contribution. The resulting expression  for
$\varepsilon(k;s)$ becomes \be
 \varepsilon(k;s)=\varepsilon(k;0)+is\,\frac{4\pi G}{k^2}\sum_a g_a\, m^3_a  \int
\frac{d^3p}{(2\pi)^3}\frac{  \frac{df_a^{(0)}(p)}{p\,dp} }{
 \frac{\vk\cdot\vp}{m_a}-is} \label{varsus}\ee where \be
 \varepsilon(k;0) = 1+\frac{4\pi G}{k^2}\sum_a  g_a\,m^3_a  \int_0^\infty p
 \frac{df_a^{(0)}(p)}{dp} \frac{dp}{2\pi^2} \equiv
 1-\frac{k^2_{fs}}{k^2}\,.
 \label{eps0}\ee In eqn. (\ref{eps0}) we have integrated by parts and
  introduced the \emph{free streaming} momentum $k_{fs}$ given by
\be k^2_{fs} =  4\pi G \sum_a
 \rho_a^{(0)} \Big\langle \frac{1}{\vec{V}^2} \Big\rangle_a \label{Kfs}\ee
 where  \be \rho^{(0)}_a = m_a n_a^{(0)} ~~;~~ n_a^{(0)}=g_a \int
 \frac{d^3p}{(2\pi)^3} f_a^{(0)}(p) \label{rho0}\ee and \be \Big\langle \frac{1}{\vec{V}^2}
 \Big\rangle_a = \frac{g_a}{n_a^{(0)}} \int
 \frac{d^3p}{(2\pi)^3}  \frac{m^2_a}{p^2} f_a^{(0)}(p)\,,
 \label{1overVave}\ee the free streaming length is obtained as
 \be \lambda_{fs} = \frac{2\pi}{k_{fs}}\,.\label{lambdafs}\ee The expression for $k_{fs}$ can be written  as \be k^2_{fs} = \frac{\Omega_{DM}
 h^2}{(\textrm{Mpc})^2}\,\sum_a \nu_a  \Big\langle \frac{V^2_0}{\vec{V}^2}
 \Big\rangle_a~~;~~V_0 = 122.5 \,\frac{\mathrm{km}}{\mathrm{s}}
 \label{nukfs}\ee where   the \emph{partial
 fractions} are defined as \be \nu_a = \frac{\Omega_a}{\Omega_{DM}} ~~;~~\sum_a
 \nu_a =1\;. \label{pf}\ee

 The \emph{collective} modes correspond to the \emph{zeroes} of the
 ``gravitational polarization'' function $\varepsilon(k;s)$ eqn. (\ref{varsus}) in the
 complex $s-$ plane. These yield the  time evolution  for the
 gravitational perturbations \be \varphi^{(1)}(\vk;t) \propto \sum_p
 e^{s_p(k)t} \label{collmode}\ee where $s_p(k)$ are the zeroes of
 $\varepsilon(k;s)$.

 It is illuminating to compare the expression  (\ref{Kfs}) with that
 for the Jeans wavevector for a single fluid \be k^2_{J} =\frac{ 4\pi G
 \rho^{(0)}}{c^2_s}\,, \label{KJ}\ee where $c_s$ is the adiabatic
   speed of sound of the fluid. We see that for a single collisionless
 component we can obtain the free streaming length from the Jeans length by the replacement
\be  c_s  \Rightarrow \left[\Big\langle
 \frac{1}{\vec{V}^2} \Big\rangle\right]^{-\frac{1}{2}} \label{Jfsrel}\ee
 which in general \emph{is different} from replacing $c_s$ by the
 velocity dispersion
 $ \sqrt{\langle \vec{V}^2 \rangle}$. This difference becomes important
 when the unperturbed distribution function favors small values of
 the momentum.
 This observation will become crucial when we study Bosonic
 particles and sterile neutrinos decoupled when relativistic but
 \emph{out of LTE}. As it will be seen in detail below, in these cases the distribution function favors
 the region of small momentum which leads to dramatic consequences
 in the difference between $1/\langle \vec{V}^2 \rangle$ and
 $\langle 1/\vec{V }^2 \rangle$.

 \subsection{Landau damping and Jeans instability}\label{LD}
 It is clear from (\ref{varsus}) that there is a pole in the Laplace
 transform of the Newtonian perturbation for the marginal value \be
 s=0 ~~;~~k=k_{fs} \,.\ee This is akin to the marginal value $k=k_J$
 in a fluid where $k_J$ is the Jeans wave vector, in a fluid for
 $k>k_J$ pressure gradients hinder gravitational collapse and the
 perturbations are simple acoustic oscillations, for $k<k_J$
 pressure gradients cannot prevent the collapse and the
 self-gravitational fluid undergoes the Jeans instability towards
 gravitational collapse. We can study the dynamics of collective excitations in region $k \approx
 k_{fs}$ searching for zeroes in  $\varepsilon(k;s)$   for $s\approx
 0$. The second term in (\ref{varsus}) can be evaluated by
 performing the angular integral and using that $f^{(0)}_a$ only
 depends on $p=|\vp\,|$, namely \be \int^1_{-1} \frac{d(\cos\theta) }{\frac{k p \cos\theta}{m_a}-is}
= \frac{m_a}{kp} \ln\left[\frac{\frac{k p }{m_a}-is}{-\frac{k p
}{m_a}-is} \right] \,. \ee

  For $s \approx 0$ the branch is defined by  the
 prescription determined by the Bromwich contour $Re(s)>0$, which is
 precisely recognized as the Landau prescription for the evaluation
 of the integrals\cite{plasmas}. We find \be \varepsilon(k;s) =
 1-\frac{k^2_{fs}}{k^2}+ is \frac{G}{\pi k^3} \sum_a g_a\,m^4_a \int_0^{\infty}
 \left[ -\frac{df^{(0)}_a(p)}{dp} \right]\left\{\ln \left[ \frac{\frac{kp}{m_a}+ is}{\frac{kp}{m_a}- is}
 \right] -i\pi\right\}dp \label{vare} \ee

Because for small $s$ the logarithm in (\ref{vare}) is linear in
$s$,   the last term  in the bracket ($-i\pi$) contributes to  the
leading order in $k^2-k^2_{fs}$.  To lowest order in $s$ for
 $k \approx k_{fs}$, the condition $\varepsilon(k;s) =0$ yields \be
 s(k) =  \mathcal{C} \left[k^2_{fs}-k^2\right]~~;~~\mathcal{C}=\frac{k}{G \sum_a g_a\, m^4_a
 f_a^{(0)}(0)}>0 \,.
 \label{polecond} \ee From the time evolution of the gravitational perturbation eqn. (\ref{collmode}) we find \bea && s(k) < 0 ~\mathrm{for} ~k >k_{fs} \Rightarrow
 ~\mathrm{Landau ~ damping} \nonumber \\ && s(k) > 0 ~\mathrm{for} ~k <k_{fs} \Rightarrow
 ~\mathrm{Jeans ~ instability }\,. \label{sss}\eea

 The long wavelength limit $k\rightarrow 0$ is obtained by expanding
 $(\frac{\vec{k}\cdot\vec{p}}{m}-is)^{-1}$ in
  the integrand in eqn. (\ref{varsus})  in powers of  $\vec{k}\cdot\vec{p}/ms$. In the resulting expression
  only the odd powers survive the angular integration.  Keeping up to $(\vec{k}\cdot\vec{p}/ms)^3$
  the long wavelength limit of $\varepsilon(k;s)$ is found to be
 \be \varepsilon(k;s) = 1- \frac{4\pi G}{s^2}\sum_a
 \rho^{(0)}_a\left[1- \big \langle V^2 \big\rangle_a \frac{k^2}{s^2}
 +\cdots \right] \label{longwav}\ee and the zeroes of $\varepsilon(k;s)$
 in the long-wavelength limit are found to be \be s_\pm(k) = \pm \Big[\Omega^2_J -
 \overline{V^2} k^2 \Big]^{\frac{1}{2}}+\cdots \label{spm}\ee where
 \be \Omega^2_J = 4\pi G \sum_a\rho^{(0)}_a ~~;~~\overline{V^2} =
 \sum_a \nu_a \big \langle \vec{V}^2 \big\rangle_a \label{pars}\ee and
 $\nu_a$ are the partial fractions. The Jeans frequency $\Omega_J$
 is the same as that for single component fluids, however the
 relationship between the free streaming wavevector $k_{fs}$ (\ref{Kfs}) and the
 Jeans frequency $\Omega_J$ is different from that of the Jeans wavevector and the Jeans frequency in a single
 component fluid.  For a single collisionless component  we find
 \be \Omega_J =
 \Bigg[\Big\langle\frac{1}{\vec{V}^2}\Big\rangle\Bigg]^{-\frac{1}{2}} \,
 k_{fs}
 \ee whereas for a single \emph{fluid} one finds\cite{peebles} \be \Omega_J = c_s k_J \,.\ee

 \subsection{An example: the Maxwell-Boltzmann distribution}
In general the momentum integral in (\ref{varepsilon}) cannot be
found in closed form without approximating the distribution
function. However the Maxwell-Boltzmann distribution provides an
example for which a closed form expression for (\ref{varepsilon})
can be found. This distribution function is relevant for the
description of WIMPs which are heavy relics that decoupled in LTE
while non-relativistic\cite{kt}. In this case \be f^{(0)}(\vec{p}) =
\mathcal{N} \, e^{-\frac{\vec{p}^2}{2mT_d}} \label{maxbol}\ee where
the normalization $\mathcal{N}$ is obtained from the solution of the
kinetic equation for the distribution function, which can be found
in section (5.2) in ref.\cite{kt}. The particle density is \be \rho
= m g \int \frac{d^3p}{(2\pi)^3}\,f^{(0)}(\vec{p})= \mathcal{N} m g
\Big[\frac{m T_d}{2\pi} \Big]^\frac{3}{2} \,. \label{MBrho}\ee The
momentum integrals in eqn. (\ref{varepsilon}) can be carried out
straightforwardly because $f^{(0)}(\vec{p})$ is a function of
$\vec{p}^2$. This is achieved by splitting the vector $\vec{p}$ into
components parallel and perpendicular to $\vec{k}$. With
$d^3p=dp_{\parallel}\,d^2p_{\perp}$ the integrals along the parallel
and perpendicular directions can be done straightforwardly. We
obtain the result \be \varepsilon(k;s) = 1- \frac{k^2_{fs}}{k^2} +
\frac{s}{k^3} \, g\mathcal{N}m^4\,
e^{\delta^2}\,\Bigg[1-\frac{2}{\sqrt{\pi}}\,\int^\delta_0
e^{-t^2}dt\Bigg]~~;~~\delta = \frac{s}{k}\,\Big[\frac{m}{2T_d}
\Big]^\frac{1}{2}\,. \label{epsiMB}\ee It is straightforward to
analytically continue this function   to $\mathrm{Re}(s) < 0$
according to the Landau prescription\cite{plasmas}.

With $f^{(0)}(\vec{0}) = \mathcal{N}$ we find for the marginal case
$s=0$ the solution  $k^2 = k^2_{fs}$,  where \be k^2_{fs} = 4\pi G
\rho \, \Big\langle \frac{1}{\vec{V}^2} \Big\rangle \,, \ee and
$\rho$ is given by eqn. (\ref{MBrho}). Furthermore it is a simple
exercise to confirm that  near the marginal case the pole in the
function $\varepsilon(k;s)$ is given by eqn. (\ref{polecond}). For
the Maxwell-Boltzmann distribution it follows that \be \Big\langle
\vec{V}^2 \Big\rangle = \frac{3T}{m} ~~;~~ \Big\langle
\frac{1}{\vec{V}^2} \Big\rangle = \frac{m}{T} \neq
\frac{1}{\Big\langle \vec{V}^2 \Big\rangle }\,. \label{VMB}\ee

The long-wavelength limit can be obtained   by expanding the
integral in (\ref{epsiMB}) for $\delta \gg 1$, namely \be
\Bigg[1-\frac{2}{\sqrt{\pi}}\,\int^\delta_0 e^{-t^2}dt\Bigg] =
\frac{e^{-\delta^2}}{\delta \,\sqrt{\pi}}\Big[1-
\frac{1}{2\,\delta^2} + \frac{3}{4\,\delta^4} + \cdots\Big]
\label{expaMB}\ee which after tedious but straightforward algebra
leads to the expressions for the poles given by eqn. (\ref{pars})
with $\nu=1$ and  $\Omega^2_J = 4\pi G \rho$  where $\rho$ is given
by (\ref{MBrho}) and $\big\langle \vec{V}^2\big\rangle$ given by
(\ref{VMB}).

 \section{Free streaming lengths for decoupled
 particles}\label{sec:deco}
The distribution function of decoupled particles in a homogeneous
and isotropic cosmological background in absence of gravitational
perturbations is constant along geodesics and obey the Liouville or
collisionless Boltzmann equation in terms of an affine parameter
$\lambda$\cite{bernstein,kt,coldmatter} \be \frac{d}{d\lambda}
f[P_f;t] =0 \Rightarrow \frac{d f[P_f;t]}{dt}= 0 \label{lio}\ee
where   $P_f = p_c/a(t)$ is the physical momentum, and $p_c$ the
time independent comoving momentum. Taking $P_f$ as an independent
variable this equation leads to the familiar
form\cite{bernstein,kt,coldmatter} \be \frac{\partial
f[P_f;t]}{\partial t}- H P_f \frac{\partial f[P_f;t]}{\partial P_f}
= 0 \,,\label{maslio}\ee where $H=\dot{a}/a$ is the Hubble parameter
and $P_f a=p_c=\mathrm{constant}$ is a characteristic of the
equation. Obviously a solution of this equation is \be f[P_f;t]
\equiv f_d[a(t)P_f] = f_d[p_c]\,. \label{fD}\ee  If a particle of
mass $m$ has been in LTE but it decoupled from the plasma with
decoupling temperature $T_d$ its distribution function is \be
f_d(p_c) = \frac{1}{e^{\frac{\sqrt{m^2+p^2_c}-\mu_d}{T_d}}\pm 1}
\label{LTEdist}\ee for Fermions $(+)$ or Bosons $(-)$ respectively
allowing for a chemical potential.

 Since the distribution function is dimensionless,
without loss of generality we can always write   for a
\emph{decoupled} particle\cite{coldmatter}    \be f_d(p_c)=
f_d\left(\frac{p_c}{T_d};\frac{m}{T_d};\alpha_i\right)
\label{farbi}\ee where  $\alpha_i$ are a collection of
\emph{dimensionless} constants determined by the microphysics, for
example dimensionless couplings or ratios between $T_d$ and particle
physics scales or   in equilibrium $\mu_d/T_d$ etc.  To simplify
notation in what follows we will not include explicitly the set of
dimensionless constants $m/T_d;\alpha_i$, etc,  in the argument of
$f_d$, but these are implicit in generic distribution functions. If
the particle decouples when it is still relativistic $m/T_d
\rightarrow 0$.

It is convenient   to introduce the dimensionless
ratios\cite{coldmatter} \be y = \frac{p_c}{T_d}=
\frac{P_f}{T_d(t)}~~;~~T_d(t)=\frac{T_d}{a(t)}~~;~~ x_d=
\frac{m}{T_d}\,. \label{yvar}\ee   We emphasize that the
distribution
 functions  (\ref{farbi}) are general and  \emph{not} necessarily  describing particles decoupled while in local thermal
 equilibrium. When the particle becomes non-relativistic, its
 contribution to the energy density is \be \rho = m\, n(t)
 \label{rho}\ee where\cite{coldmatter} \be n(t)=g~
\frac{T^3_d(t)}{2\pi^2}\int^\infty_0 y^2 f_d(y) dy  \label{nnoft}\ee
and $g$ is the number of internal degrees of freedom.
  From entropy conservation\cite{kt,bernstein}, the decoupling temperature at   redshift $z$   is
 related to the temperature of the CMB today  by \be T_d(z) = (1+z)
 \left(\frac{2}{g_d}\right)^{\frac{1}{3}}T_{cmb}= \left(\frac{2}{g_d}\right)^{\frac{1}{3}} 2.348\times
 10^{-4}(1+z)
 ~\textrm{eV}\;,\label{Tdo}\ee where $g_d$ is the number of effective ultrarelativistic degrees of freedom
 at decoupling.  For a given species $(a)$  of   particles with $g_a$ internal degrees of
freedom that decouples when the effective number of
ultrarelativistic degrees of freedom is $g_{d,a}$, the  relic
abundance \emph{today} ($z=0$)  is given by\cite{coldmatter} \be
\Omega_a \,h^2 = \left(\frac{m_a}{25.67~\mathrm{eV}}\right)
\frac{g_a\int^\infty_0 y^2 f_{d,a}(y)dy}{2 g_{d,a}\,\zeta(3)} \,.
\label{relic} \ee  If this decoupled species contributes a fraction
$\nu_a$ to dark matter, with $\Omega_a = \nu_a \Omega_{DM}$  and
taking $\Omega_{DM}h^2 = 0.105$\cite{wmap3} for non-baryonic dark
matter, we find\cite{coldmatter} \be \nu_a =
\left(\frac{m_a}{6.227~\mathrm{eV}}\right)
\frac{g_a}{g_{d,a}}\int^\infty_0 y^2 f_{d,a}(y)dy \,. \label{nua}\ee
The constraint on the    fractional abundance   $0\leq \nu_a \leq 1$
yields a bound for the mass of the particle\cite{coldmatter}.

 We are now in position to establish a relation between the
 distribution function of the  decoupled particles as dark matter candidates and the results from
 the Boltzmann-Vlasov equation  which determine the   free streaming length.

 The analysis of the Boltzmann-Vlasov equation in the non-expanding
 case
 applies to the description of perturbations at very low redshift in
 the Newtonian approximation. Such approximation is correct provided
 the relevant length scales, for example the free streaming length,
 are much smaller than the Hubble radius. Such is the case whenever the velocity dispersion of
 the particles is $(V/c)^2 \ll 1$. The momentum that enters in the
 Boltzmann-Vlasov equation is the physical momentum, which  in the non-relativistic
 limit is related to the velocity as
 \be \vec{V}^2 = \frac{\vec{P}^2_f}{m^2}\,. \label{Vnrel}\ee The   unperturbed distribution
 functions for decoupled particles that enter in the linearized Boltzmann-Vlasov eqn. (\ref{Vlasopert})
  are the solutions of the unperturbed collisionless Boltzmann
 equation, namely \be f^{(0)}(p)=f^{(0)}_d(p_c)\label{fdec}\ee where $f_d(p_c)$
 are given by eqn. (\ref{farbi}).   Restoring the speed of light $c$,
the average \be \Big\langle \frac{1}{\vec{V}^2}
 \Big\rangle_a = \left(\frac{m_a}{T_{d,a}(z)\,c}\right)^2 I_a  \,,
 \label{1oV2}\ee where $T_{d,a}(z)$ is given by eqn. (\ref{Tdo}) for the species $a$, and
 we have introduced  \be I_a \equiv \Bigg[
 \frac{\int_0^\infty f^{(0)}_{d,a}(y)dy}{\int_0^\infty
 y^2\,f^{(0)}_{d,a}(y)dy}\Bigg]\,.\label{Ia}\ee This dimensionless ratio of integrals only depends on the
 ratios $m_a/T_d,\mu_a/T_d$ and dimensionless couplings from the
 microphysics at decoupling.  Using eqn. (\ref{Tdo}) we obtain   \be \Big\langle \frac{V^2_0}{\vec{V}^2}
 \Big\rangle_a =  \frac{1.905}{(1+z)^2}\left(\frac{m_a}{\mathrm{eV}}\right)^2 \,
 g^{\frac{2}{3}}_{d,a}\,I_a\,.
 \label{V0oV2} \ee Inserting this result into the expression (\ref{nukfs}) and using $\Omega_{DM}h^2 = 0.105$\cite{wmap3}
  for non-baryonic dark
matter we find  \emph{today}, for $z=0$ \be k^2_{fs} =
 \Big[\frac{0.447}{\textrm{kpc}}\Big]^2 \, \sum_a \nu_a\,g^\frac{2}{3}_{d,a}
 \left(\frac{m_a}{\mathrm{keV}}\right)^2 \,I_a \,.
 \label{kfs2}\ee

 For light particles that decouple while they are
 ultrarelativistic the distribution function $f_{d,a}(y)$ does not
 depend on the ratio $m_a/T_d$, however, for particles that decouple when they are non-relativistic, their
 distribution function is typically a Maxwell-Boltzmann distribution which does depend on this
 ratio.

 Inserting the result (\ref{nua}) for the   fractions (\ref{nua})
 and (\ref{V0oV2}) into eqn. (\ref{kfs2}) leads to the alternative form
  \be k^2_{fs} =   \Big[\frac{  5.632}{\textrm{kpc}}\Big]^2 \, \sum_a
  \frac{g_a}{g^{\frac{1}{3}}_{d,a}}
 \Big(\frac{m_a}{\mathrm{keV}}\Big)^3 \int_0^\infty f^{(0)}_{d,a}(y) dy
   \label{kfsfin}\ee where $g_a$ are the internal degrees of freedom
   of the particle of species $(a)$. The  simple
   expressions (\ref{kfs2},\ref{kfsfin}) are some of  the main results of this article.

\subsection{Redshift dependence}\label{subsec:redshift}

 The  Boltzmann-Vlasov equation in a
 non-expanding cosmology (\ref{vlasov}) is obtained by setting the scale factor $a \equiv 1$ in
 the same equation  in an expanding cosmology (see
 ref.\cite{peebles,bert}). Thus the linearized equation describes the
 properties of the  collective excitations  \emph{today}. In an
 expanding cosmology the decay or growth of perturbations is no
 longer exponential but typically a power   of the scale
 factor\cite{peebles,bert}. However, here we are \emph{not} concerned directly
 with the manner in which linear perturbations grow or decay, but
 with the marginal wave-vector $k_{fs}$ that determines the
 crossover of behavior from growth to damping of collective
 excitations. If  we \emph{assume} that the expansion is
 sufficiently slow that it can be treated adiabatically we can obtain the redshift
 behavior of the free-streaming length $\lambda_{fs}(z)$  by replacing
 the densities, velocities and wavevectors with the corresponding
 scale factors given by eqns. (\ref{nnoft},\ref{Tdo},\ref{V0oV2}), namely: \bea   \rho & \rightarrow & \rho(z) = \rho( 0)(1+z)^3 \nonumber \\
 \Big\langle \frac{1}{\vec{V}^2}\Big\rangle  & \rightarrow & \Big\langle \frac{1}{\vec{V}^2 (z) } \Big\rangle  =
 \Big\langle \frac{1}{\vec{V}^2 (0) }\Big\rangle
 (1+z)^{-2} \nonumber \\
 k  & \rightarrow &   k   (1+z) \,,
\label{redshifted} \eea  where $z=0$ refers to \emph{today}.
Defining the \emph{comoving} free-streaming wavevector $k_{fs}(z)=
2\pi /\lambda_{fs}(z)$ upon rescaling by the corresponding scale
factor \be k_{fs} \rightarrow k_{fs}(z)\,(1+z)\, .
\label{kfsscale}\ee
 Assuming the validity of  this adiabatic scaling in eqn (\ref{Kfs}) we
obtain, \be
 k^2_{fs}(z)  = \frac{4\pi G}{(1+z)^2} \sum_a
\rho^{(0)}(z)~\Big\langle \frac{1}{\vec{V}^2(z)}\Big\rangle \,.
\label{soluscal}\ee Since the DM density and velocity dispersions
for all components scale as in eqn. (\ref{redshifted}) we find the
following redshift dependence of the \emph{comoving} free-streaming
wavevector \be k^2_{fs}(z) = \frac{4\pi G}{(1+z)} \sum_a \rho^{(0)}(
0)~\Big\langle \frac{1}{\vec{V}^2 (0) }\Big\rangle \,.
\label{soluscalfin}\ee

This result is   similar to the expression for the Jeans's
wavevector in a Newtonian fluid\cite{peebles,bert} upon replacing
$\big\langle {1}/{\vec{V}^2(z)}\big\rangle \rightarrow c^2_s(z)$
where $c_s(z)$ is  the (adiabatic) speed of sound in the medium as a
function of redshift.  In fact
 the validity of this assumption is confirmed not only by the similarity with the
 familiar  Jeans'  result for Newtonian fluids in an expanding cosmology, but also by the exact solution
 obtained in ref.\cite{ringwald} for the case of neutrinos decoupled
 in LTE. Therefore we identify $k_{fs}$ given by eqn. (\ref{kfs2}) as the \emph{comoving}   free streaming
 wave-vector. The scaling behavior of the \emph{comoving} free streaming length
 $\lambda_{fs}(z) =  \lambda_{fs}(0) \sqrt{(1+z)} $ leads to the free-streaming
 mass \be M_{fs}(z) =  \frac{4\pi}{3} \sum_a
 \rho^{(0)}_a(z)\, \Bigg(\frac{\lambda_{fs}(z)}{ 1+z }\Bigg)^3   = M_{fs}(0) (1+z)^{\frac{3}{2}}
 \label{Mfs}\,, \ee a relation similar to the that of the Jeans'
 mass in the non-relativistic regime. Therefore, under the
 validity of the adiabatic assumption, the simple re-scaling of the
 free-streaming wave-vector and length given by eqn. (\ref{soluscalfin})
 indicates that we can simply obtain these quantities \emph{today}
 ($z=0$) and extrapolate to an arbitrary redshift $z$ via eqn.
 (\ref{soluscalfin}) provided the redshift is still small enough that the species are
 non-relativistic.

 The validity of the adiabatic assumption relies on the fact that in
 the non-relativistic regime with $\langle \vec{V}^2/c^2 \rangle \ll
 1$, the free-streaming length is \emph{much smaller} than the
 Hubble radius, which is found below to be a consistent assumption,
 or alternatively $  {k_{fs}}/{H} \gg 1 $. And as mentioned above the result for the
 free streaming length obtained from this adiabatic hypothesis is
 similar to the usual result for the Jeans' length\cite{peebles,bert} and
 is confirmed in ref.\cite{ringwald} for the case of a neutrino
 thermal relic.

 A more detailed analysis of Gilbert's equation for \emph{mixtures} of DM
 components with arbitrary (but isotropic) distribution functions in the adiabatic approximation will be provided
 elsewhere\cite{next}.

 We now gather the above results to give the general expression for the free streaming length
 of an arbitrary \emph{mixture} of non-relativistic species that decoupled in or out
of LTE either ultrarelativistic or non-relativistic,  in terms of
the  \emph{ partial fraction} ($\nu$) that each contributes to the
(DM) content and the dimensionless ratios $I_a$   \be
\frac{1}{\lambda^2_{fs}(z)} = \frac{1}{(1+z)}\,
\Big[\frac{0.071}{\textrm{kpc}}\Big]^2 \, \sum_a
\nu_a\,g^\frac{2}{3}_{d,a}
 \left(\frac{m_a}{\mathrm{keV}}\right)^2 \,I_a \,,\label{lamfsz1} \ee
 where \be I_a = \Bigg[
 \frac{\int_0^\infty f^{(0)}_{d,a}(y)dy}{\int_0^\infty
 y^2\,f^{(0)}_{d,a}(y)dy}\Bigg] \label{Ia2}\ee  in terms of the general distribution functions   (\ref{farbi})
 which   only depend on the
 ratios $m_a/T_d,\mu_a/T_d$ and dimensionless couplings and are
 completely determined by the  microphysics at decoupling.

\medskip

 We now proceed to obtain the contributions to the free streaming
 wave-vector \emph{today} ($z=0$) from the various   components:
 thermal relics that decoupled either relativistic or
 non-relativistic in LTE and non-thermal relics that decoupled while
 relativistic but out of LTE.

   \subsection{Thermal relics}
   Let us consider ultrarelativistic Fermionic or Bosonic particles
   decoupled in LTE   with chemical potentials and with $m_a/T_{d,a}
   \ll1$.

   \begin{itemize}
   \item{{\bf Ultrarelativistic Fermions:}
   Neglecting $m/T_d$ in the ultrarelativistic limit, but keeping the chemical potential $\mu$,
    the distribution function is \be f^{(0)}_{d}(y)=
   \frac{1}{e^{(y-u)}+1}~~;~~u = \frac{\mu}{T_{d}}
   \label{fermi}\ee where we have neglected $m/T_{d} \ll 1$. For this distribution
    \be \int_0^\infty f^{(0)}_{d}(y) dy =
   \ln[1+e^{u}]\label{fer} \ee Combining this result with eqn. (\ref{kfsfin}) we note that
    larger chemical potentials   lead  to \emph{shorter} free streaming scales.

   Denote $I_F[u]$ the ratio $I_a$, eqn. (\ref{Ia})  for an ultrarelativistic Fermionic thermal relic with chemical
   potential $\mu$. It is depicted in fig. (\ref{fig:fermions}) as a
   function of $u=\mu/T_d$. For $\mu=0$ we find \be I_F[0] =
   \frac{2\,\ln(2)}{3\,\zeta(3)} = 0.3844 \,.
   \label{I0}\ee

\begin{figure}[h]
\begin{center}
\includegraphics[height=3in,width=3in,keepaspectratio=true]{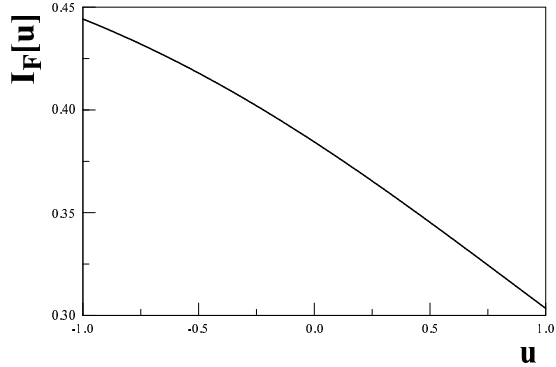}
\caption{$I_F[u]  $ vs $u=\mu /T_d$ for $f_d(y)= 1/(e^{(y-u)}+1)$.
$I_F[0]=0.3844$.} \label{fig:fermions}
\end{center}
\end{figure}

 For $\mu=0$ and taking $\nu=1$ for this Fermionic species and vanishing abundance for all others, the result
 for $k_{fs}$ given by eqn. (\ref{kfs2}) with $I_a =  {2\,\ln(2)}/{3\,\zeta(3)}$ agrees with that found
 in ref.\cite{ringwald}.  Therefore, a Fermionic thermal relic as a unique DM component yields a free-streaming
 length \emph{today} ($z=0$)
 \be \lambda_{fs}(0) = \frac{14}{g^\frac{1}{3}_d\,\sqrt{I_F[u]}}  \,\Big(
 \frac{\mathrm{keV}}{m}\Big)\,   \mathrm{kpc}\,.
 \label{lfsfer}\ee  For example  a
  neutrino with $m \sim \mathrm{eV}$ decoupling at $T_d \sim 1\,\mathrm{MeV
 }$, when $g_d \sim 10$, yields a free streaming length today \be
 \lambda_{fs}(0) \sim 10\,\mathrm{Mpc}\,, \ee which is the usual estimate for free-streaming lengths for
 HDM candidates. }

 \item{{\bf Ultrarelativistic Bosons:} The distribution function is \be f^{(0)}_{d}(y)=
   \frac{1}{e^{(y-u_d)}-1}~~;~~u_d  = \frac{\mu }{T_{d}}
   \label{bose}\ee where we have neglected $m/T_{d} \ll 1$ and in this case the chemical
   potential $\mu \leq 0$ so that the distribution function (a probability) is manifestly
   positive definite. For this distribution
    \be \int_0^\infty f^{(0)}_{d}(y) dy = -
   \ln[1-e^{-|u_d|}] \label{bos}\ee This expression clearly reveals that  for $m=0,\mu_d=0$
    the distribution function diverges logarithmically at $p=0$. In
    an expanding cosmology with a particle horizon the smallest
    wavevector that can describe causal microscopic physics
      is determined by the Hubble scale. Since we are considering decoupling in LTE,
    it is consistent to assume an infrared momentum cutoff   of order
    $H_d \sim \sqrt{g_d}\, T^2_d/M_{Pl}$, the Hubble parameter at the time of decoupling. Therefore
 \emph{assuming} such an infrared cutoff, keeping $m/T_d \ll 1$ but non-zero and with vanishing
 chemical potential we find  \be \int_0^\infty f^{(0)}_{d}(y) dy
 \sim \ln\Bigg[\frac{2\,T_d}{H_d+\sqrt{H^2_d+m^2}}\Bigg]\,. \label{URBE}
 \ee In the denominator in eqn. (\ref{Ia}) we can set $x_d=0$ in the ultrarelativistic limit and for
 vanishing chemical potential we find \be I_B[x_d,0] \sim
 \frac{1}{2\zeta(3)}\, \ln\Bigg[\frac{2\,T_d}{H_d+\sqrt{H^2_d+m^2}}\Bigg]\,.
 \label{IBnomu}\ee

     Keeping the mass
    and the chemical potential the distribution function becomes
     \be f^{(0)}_{d}(y) = \frac{1}{e^{\sqrt{y^2+x^2_d}-u_d}-1}
    ~~;~~x_d = \frac{m}{T_d} ~~;~~u_d = \frac{\mu}{T_d}\ee
     for which we find
     \be \int_0^\infty f^{(0)}_{d}(y) dy = x_d\, \sum_{l=1}^\infty
     e^{-l|u_d|}\,K_1\big[l\,x_d\big]\,, \label{fint}\ee Where $K_1$ is a Hankel function. For $m/T_d,
     \mu/T_d \ll 1$ we can neglect the mass and chemical potential
     in  the denominator in eqn. (\ref{Ia}) and we find for an
     ultrarelativistic boson with a non-vanishing chemical potential
     in the limit $m/T_d,
     \mu/T_d \ll 1$ \be I_B[x_d;u_d] \simeq  \frac{x_d}{2\zeta(3)} \sum_{l=1}^\infty
     e^{-l |u_d|}\,K_1\big[ l\,x_d \big]  \,.
     \label{Ib}\ee This function features  an infrared divergence in
     the limit $m/T_d,\mu/T_d \rightarrow 0$ from the numerator in
     (\ref{Ia}) given by (\ref{fint}), which must be regulated by assuming an infrared cutoff of
     the order of the Hubble scale at decoupling as in eqn.
     (\ref{URBE}). A more thorough analysis of the infrared behavior
     of the decoupled distribution function in an expanding cosmology is
     necessary, in particular an assessment of the (causal) thermalization
     process for superhorizon wavelengths. Such   program,
     interesting all by itself, is clearly beyond the realm of this
     study.

     {\emph{Bose-Einstein Condensation}}: If chemical freeze out
     occurs before kinetic decoupling, it is possible that bosonic
     particles undergo Bose-Einstein condensation\cite{coldmatter}. Under this
     circumstance the homogeneous condensate ($\vec{p}=0$)
     contributes to the dynamics of the scale factor and
     inhomogeneous perturbations only modify the distribution
     function of the particles outside of the condensate. The
     linearized Boltzmann-Vlasov equation therefore applies to the
     non-condensate part, since the dynamics of the condensate
     cannot be described by the incoherent Boltzmann equation but by
     the equation of motion of the coherent and homogeneous
     condensate, which in turn is coupled to the Friedmann equations
     for the scale factor.

     In the case of Bose-Einstein condensation the chemical
     potential attains its maximum value $\mu = m$ in which case the
     contribution from the particles \emph{outside} the condensate
     is  $I_B[x_d,x_d]$ which has a strong infrared divergence \emph{for any value
     of the mass}. This is a consequence of the fact that for
     ultrarelativistic Bose-Condensed particles the distribution
     function diverges at $p=0$ \emph{for any value of the mass}.

     However, as discussed above  this  infrared divergence associated with a Bose-Einstein condensate
     must be carefully assessed in an expanding cosmology
     because infrared modes with wavelengths larger than the size of the horizon
   will not be  in LTE since no causal processes can establish thermalization for superhorizon wavelengths.
   Hence we \emph{conjecture} that the integrals of the Bose-Einstein distribution function must be cutoff in the infrared
   at a momentum of the order of the Hubble scale at decoupling, $H_d \sim \sqrt{g_d}\, T^2_d/M_{Pl}$.
    Such a cutoff   leads to the estimate
    \be I_{BEC} \sim   \frac{m\,M_{Pl}}{\sqrt{g_d}~T^2_d}\sim \frac{10^{12} }{\sqrt{g_d}}\,
    \Big(\frac{m}{\mathrm{keV}} \Big)\,\Big( \frac{\mathrm{GeV}}{T_d}\Big)^2 \label{cutoff}\,.
    \ee If this were the \emph{only} DM component, the resulting
    comoving free-streaming length at $z=0$ (\emph{today}) is given by \be \lambda_{fs}(0) \sim
    \frac{0.014\,\mathrm{pc}}{g^\frac{1}{12}_d}\,\Big(\frac{\mathrm{keV}}{m} \Big)^\frac{3}{2}\,
    \Big(\frac{T_d}{\mathrm{GeV}} \Big)  \ee

    It is clear from the discussion above that the microphysics of
    decoupling of light bosonic particles, with or without a
    Bose-Einstein condensate requires a thorough assessment of
    the infrared behavior of the distribution function. The primordial velocity dispersion is
    very sensitive to this cutoff whose origin lies in the causal aspects of decoupling.
    This important physical aspect
   must be studied in deeper detail, a task that is certainly beyond the realm of this article, but we
   can nevertheless conclude that a light bosonic particle decoupled in LTE can effectively act as
   CDM as a consequence of the infrared sensitivity of the moment
   $\langle 1/\vec{V}^2 \rangle$ for Bosonic thermal relics either condensed or not.   }

   \item{{\bf Non-relativistic particles:} The distribution function
   after freeze out in LTE  is the Maxwell-Boltzmann distribution \be f^{(0)}_d(p_c) = n_d
   \left[\frac{m T_d}{2\pi}\right]^{-\frac{3}{2}}
   e^{-\frac{\vp^2_c}{2mT_d}}=  n_d
   \left[\frac{m T_d}{2\pi}\right]^{-\frac{3}{2}}e^{-  \frac{y^2 T_d}{2m}}\label{MB} \ee where $n_d$
    is the number of  particles per comoving volume at freeze-out\cite{kt} \be n_d = \frac{2\pi^2}{45} g_d T^3_d \, Y_\infty\ee
    and $Y_\infty$ is obtained from
   the solution of the kinetic equation and is a function of  the annihilation cross section (see section 5.2 in
   ref.\cite{kt}). We
   find \be \int_0^\infty f^{(0)}_d(y)dy = \frac{4\pi^4 \,g_d\,Y_\infty}{45\,x_d} \ee and for the integral
   $I_a$ eqn. (\ref{Ia})
   denoted by $I_{NR}$ for the non-relativistic (Maxwell-Boltzmann) distribution we obtain \be I_{NR} =
   \frac{T_d}{m}\,.
   \label{INR}\ee In the case of (WIMPs)   with\cite{dominik}, $ m \sim 100 \,\mathrm{GeV}~;~T_d \sim
   10\,\mathrm{MeV}$ as candidates  for cold dark matter,  $x_d\sim 10^{-4}$. If this is the \emph{only}
    DM candidate with $\nu = 1$ with
   vanishing abundance of the other WDM or HDM candidates, the comoving free streaming length at $z=0$
   is given by
   \be \lambda_{fs}(0) \sim \frac{0.014\,\mathrm{pc}}{g^\frac{1}{3}_d}\,\Big( \frac{100\,\mathrm{GeV}}{m}\Big)^\frac{1}{2}\,
   \Big( \frac{10\,\mathrm{MeV}}{ T_d}\Big)^\frac{1}{2}\,, \ee from which it follows that for WIMPs with $m \sim
   100\,\mathrm{GeV}$ that decouple kinetically at $T_d \sim 10
   \,\mathrm{MeV}$\cite{dominik} when $g_d \sim 10$\cite{kt}  \be \lambda_{fs}(0) \sim   6.5\times
   10^{-3}\,\mathrm{pc}\,.\label{wimplam}\ee}

\end{itemize}

\subsection{Decoupling out of LTE} In ref.\cite{coldmatter}
 the following  distribution function for particles that decouple \emph{out
of LTE} and that effectively models several cases of cosmological
relevance was introduced,
 \be f_d(y) = f_0 f_{eq}\Big(\frac{y}{\eta}\Big)\, \theta(y_0-y)\,,
\label{cascade}\ee where $f_{eq}\big(\frac{p_c}{\eta T_d}\big)$ is
the equilibrium distribution function for a relativistic particle at
an effective temperature $\eta T_d$. This form  is motivated by
detailed studies of production\cite{boydata} and  thermalization
process that proceeds by energy transfer from long to short
wavelengths via a cascade with a \emph{front} that moves towards the
ultraviolet\cite{dvd}. If the interaction rate for mode mixing
becomes smaller than the expansion rate the advance of this front is
\emph{interrupted} at a fixed value of the momentum, identified here
to be $p^0_c= y_0 T_d$ where $T_d$ is the temperature of the
environmental degrees of freedom that are in LTE at the time of
decoupling\cite{coldmatter}.   The amplitude $f_0$ and effective
temperature $\eta T_d \leq T_d$ reflect an incomplete thermalization
behind the front of the cascade and determine the average number of
particles in its \emph{wake}\cite{dvd}. The non-equilibrium
distribution function (\ref{cascade}) yields a fairly accurate
description of these processes and the decoupling out of LTE.

\medskip

\noindent Remarkably, this non-LTE distribution function also
describes\cite{coldmatter} sterile neutrinos produced non-resonantly
via the Dodelson-Widrow\cite{dw} (DW) mechanism or resonantly via a
lepton-driven MSW resonance\cite{este}.

\begin{figure}[h]
\begin{center}
\includegraphics[height=3in,width=3in,keepaspectratio=true]{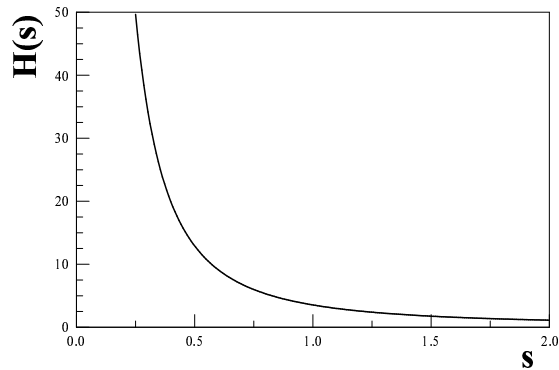}
\caption{$H(s)  $ vs $s$ for $f_{eq}(y)= 1/(e^y+1)$ .}
\label{fig:hofs}
\end{center}
\end{figure}

For the general form (\ref{cascade})  of the distribution function
we find \be I_a = \frac{1}{\eta^2}\, H\Big[
\frac{p^0_c}{\eta\,T_d}\Big] ~~;~~ H[s] = \frac{\int^s_0
f_{eq}(y)dy}{\int^s_0 y^2 f_{eq}(y)dy}\,. \label{INLTE} \ee The
function $H(s)$ for $f_{eq}(y)= 1/(e^y+1)$  is a monotonically
decreasing function of $s $ with limiting behavior $H(s) \sim 3/ s^2
~ ~ \textrm{for}~s\rightarrow 0$ and $H(s) \rightarrow
2\ln(2)/3\zeta(3)~~\textrm{for}~s \rightarrow \infty$, it is
displayed in fig. (\ref{fig:hofs})  in the interval $0.25\leq s \leq
2$.

\medskip

Although a more detailed assessment of the production and pathway
towards thermalization of sterile neutrinos is still being
debated\cite{boyho,shapo} we will adopt the semi-phenomenological
results of refs.\cite{dw,este} as   guiding models for the
distribution functions of sterile neutrinos decoupled out of LTE.

\begin{itemize}
\item{\textbf{Sterile neutrinos produced via the (DW)  mechanism}\cite{dw}: for this case
$\eta=1; s\rightarrow \infty;f_0 \sim 0.043
[\textrm{keV}/m]$\cite{dw,coldmatter}. The integral $I_a$ does not
depend on the amplitude $f_0$ (only the fractional abundance $\nu$
is affected by $f_0$) and is therefore given by the value for an
ultrarelativistic fermion with vanishing chemical potential
decoupled in LTE\cite{dw}, namely $I_{DW} = 2\ln(2)/3\zeta(3)\sim
0.3844$. \emph{If} these sterile neutrinos are the \emph{only} DM
component, namely $\nu=1$, the free streaming length today is given
by the same form as for a Fermionic thermal relic eqn.
(\ref{lfsfer}) but with vanishing chemical potential,   \be
\lambda_{fs}(0) = \frac{22.7}{g^\frac{1}{3}_d }  \,\Big(
 \frac{\mathrm{keV}}{m}\Big)\,   \mathrm{kpc}\,.
 \label{dwsterlfs}\ee  In the Dodelson-Widrow scenario\cite{dw} the sterile neutrino production
 rate peaks at $T\sim 130 \, \mathrm{MeV}$ which is near the QCD scale.
 Taking
 the decoupling temperature in this range results in $g_d \sim 30$\cite{kt}, leading to  \be \lambda_{fs}(0)
 \simeq
 {7} \, \mathrm{kpc}\,  \,\Big(
 \frac{\mathrm{keV}}{m}\Big)\,   .
 \label{dwsterlfsf}\ee Therefore $m\sim \mathrm{keV}$ sterile neutrinos produced via the (DW)
 mechanism yield free streaming lengths today of the order of $\lambda_{fs}\sim 7 \,\mathrm{kpc}$. As discussed in
 ref.\cite{dw} there is a potential ambiguity in $g_d$ because near the QCD phase transition there is an abrupt
 change in the effective number of relativistic degrees of freedom. However, because the cube-root of $g_d$
 enters
 in $\lambda_{fs}$ the free streaming length is not very sensitive
 to this ambiguity.}

\item{ \textbf{Lepton-driven resonantly produced sterile neutrinos}\cite{este}:
for this case $\eta =1, s \sim 0.7; f_0 =1$\cite{este,coldmatter},
and we find $I_a= H(0.7) = 6.814$. If this species is the
\emph{only} DM component we find the free-streaming scale
\emph{today} \be \lambda_{fs}(0) = \frac{5.4}{g^\frac{1}{3}_d }
\,\Big(
 \frac{\mathrm{keV}}{m}\Big)\,   \mathrm{kpc}\,.
 \label{esterlfs}\ee The production rate in the resonant case also
 seems to peak near the QCD scale\cite{este} at which $g_d \sim 30$. Taking
 this value for the decoupling temperature we obtain the estimate \be
\lambda_{fs}(0) =  {1.73} \,   \mathrm{kpc}\,  \,\Big(
 \frac{\mathrm{keV}}{m}\Big) .
 \label{esterlfs}\ee Hence  $m\sim \mathrm{keV}$ sterile neutrinos produced via the
 lepton-driven resonant mechanism\cite{este}
 mechanism yield free streaming lengths today of the order of $\lambda_{fs}\sim 2
 \,\mathrm{kpc}$, which is consistent with the values $\sim 3
 \,\mathrm{kpc}$ used in\cite{gao} to study the first stars forming in filamentary structures.

The enhancement of $I_a$ and consequently the shortening of the free
streaming scale in the case of sterile neutrinos produced resonantly
out of LTE follows from the fact that
    the distribution function resulting from the resonant production mechanism favors low momenta.
     Therefore for the same value of the
  contribution of sterile neutrinos produced resonantly or non-resonantly out of LTE, the free streaming
  length in the case of resonance production  is $\sim 4$ times \emph{smaller} than
  either a thermal Fermionic relic with vanishing chemical potential
   or a sterile neutrino produced non-resonantly via the (DW) mechanism.  Just as in the (DW) scenario,
   there is an ambiguity in the precise determination of $g_d$ near the QCD scale, but again the free streaming
   length is not very sensitive to this ambiguity because of the power $1/3$.  }

\end{itemize}

Although the details of the mechanism of production and decoupling
of sterile neutrinos are still under scrutiny\cite{boyho}, we take
the above results as a guideline. In particular the lesson from the
resonant production case highlights the potentially dramatic
reduction of the free streaming length when the non-equilibrium
distribution function favors smaller values of the momentum. This
observation clearly calls for a deeper assessment of the production
and thermalization of sterile neutrinos to extract reliably their
non-equilibrium distribution function.

\medskip

Combining all the results above into eqns.
(\ref{lamfsz1},\ref{Ia2}), the free streaming length for a
\emph{mixture} of Fermionic and Bosonic thermal relics, including a
possible BEC, WIMPs and sterile neutrinos produced non-resonantly or
resonantly   assumed to be described by the respective distribution
functions quoted in refs.\cite{dw,este} is given by \bea
\frac{1}{\lambda^2_{fs}(z)} = \frac{1}{(1+z)}\,
\Big[\frac{0.071}{\textrm{kpc}}\Big]^2 \sum_{\textrm{
species}}\Bigg\{&&\nu_F \,
g^\frac{2}{3}_{d,F}\,\left(\frac{m_F}{\mathrm{keV}}\right)^2 I_F[u]
+ \nu_s \,
g^\frac{2}{3}_{d,s}\,\left(\frac{m_s}{\mathrm{keV}}\right)^2\, 6.814
+\nu_B \,
g^\frac{2}{3}_{d,B}\,\left(\frac{m_B}{\mathrm{keV}}\right)^2
I_B[x_d,u_d]+ \nonumber \\&& 10^{12}\, \nu_{wimp}\, \,
g^\frac{2}{3}_{d,wimp}\,
\Big(\frac{m_{wimp}}{100\,\textrm{GeV}}\Big)\,\Big(\frac{T_{d}}{10
\,\textrm{MeV}}\Big) \Bigg\}\,,\label{kfs2tot2}\eea where the index
$F$ refers to thermal Fermions and  sterile neutrinos produced
non-resonantly via the (DW) mechanism\cite{dw} for which
$I_F[0]=2\ln(2)/3\zeta(3)=0.3844$, and the label $s$ refers solely
to sterile neutrinos produced resonantly via the mechanism in
ref.\cite{este}. In each case $g_{d,a}$ is the effective number of
ultrarelativistic degrees of freedom when the corresponding species
decouples. For sterile neutrinos produced by either mechanism $g_d
\sim 30$ corresponding to a decoupling temperature near the QCD
scale.

 \section{Conclusions and further questions}

 In this article we have implemented methods from the theory of
 multicomponent plasmas to study free streaming of a \emph{mixture} of non-relativistic
 DM candidates that include  Fermionic and Bosonic particles that
 decouple in LTE while relativistic, including the possibility of a
 Bose-Einstein condensate, heavy thermal relics that decoupled in
   when non-relativistic (WIMPs), and sterile neutrinos that
 decouple \emph{out of LTE} when they are relativistic. We solve
 \emph{exactly} the Boltzmann-Vlasov equation for the gravitational
 perturbations in a non-expanding cosmology and obtain the ``gravitational polarization function''
whose zeroes determine the dispersion relations of the collective
excitations of the self-gravitating collisionless gas of particles.
 The free-streaming wave vector is obtained from
 the marginal solution that separates Landau damped short wavelength
 perturbations from unstable collective modes. We obtain the
 free-streaming length for \emph{arbitrary} (but isotropic) distributions
 of the  particles that decoupled \emph{in or out of LTE} solely in terms of the fractional abundance
 of the different species and integrals of their distribution functions which depend on the microphysics
 at decoupling.

  Because all of the components are
non-relativistic and assuming  that the expansion is slow we provide
an adiabaticity argument  that allows us to implement a simple
rescaling of densities, velocities dispersion and wavelengths to
extract the redshift dependence of the free-streaming length. The
validity of this adiabatic approximation is confirmed by the
similarity of the result to the Jeans' length for Newtonian
perturbations in an expanding cosmology and by the explicit result
for the free-streaming wavevector for thermal neutrinos obtained in
ref.\cite{ringwald}.

 The main result for the free streaming length as a function of
 redshift for an arbitrary mixture of DM components is
\be \frac{1}{\lambda^2_{fs}(z)} = \frac{1}{(1+z)}\,
\Big[\frac{0.071}{\textrm{kpc}}\Big]^2 \,  \sum_a
\nu_a\,g^\frac{2}{3}_{d,a}
 \left(\frac{m_a}{\mathrm{keV}}\right)^2 \,I_a \,,\label{lamfsz} \ee
 where \be I_a = \Bigg[
 \frac{\int_0^\infty f^{(0)}_{d,a}(y)dy}{\int_0^\infty
 y^2\,f^{(0)}_{d,a}(y)dy}\Bigg] \label{Ial}\ee   is a ratio of integrals of the distribution
 functions that only depends on the
  $m_a/T_d,\mu_a/T_d$ and dimensionless couplings and is
 completely determined by the  microphysics at decoupling.
 Evaluating these integrals for thermal Fermionic and Bosonic relics
 (with or without condensation), WIMPs and sterile neutrinos
 decoupled out of LTE either resonantly or non-resonantly with the
 distribution functions obtained in refs.\cite{este,dw}
 respectively, we find the general result

  \bea \frac{1}{\lambda^2_{fs}(z)} = \frac{1}{(1+z)}\,
\Big[\frac{0.071}{\textrm{kpc}}\Big]^2 \sum_{\textrm{
species}}&&\Bigg\{ \nu_F \,
g^\frac{2}{3}_{d,F}\,\left(\frac{m_F}{\mathrm{keV}}\right)^2 I_F[u]
+ \nu_s \,
g^\frac{2}{3}_{d,s}\,\left(\frac{m_s}{\mathrm{keV}}\right)^2\, 6.814
+\nu_B \,
g^\frac{2}{3}_{d,B}\,\left(\frac{m_B}{\mathrm{keV}}\right)^2
I_B[x_d,u_d]+ \nonumber \\&& 10^{12}\, \nu_{wimp}\, \,
g^\frac{2}{3}_{d,wimp}\,
\Big(\frac{m_{wimp}}{100\,\textrm{GeV}}\Big)\,\Big(\frac{T_{d}}{10
\,\textrm{MeV}}\Big) \Bigg\}\,,\label{lfsconclu}\eea where $\nu_a$
is the partial fraction of each component with $\sum_a \nu_a =1$,
the sum over Fermionic species $(F)$ includes thermal relics
\emph{and} sterile neutrinos produced non-resonantly via the (DW)
mechanism\cite{dw}, for which the chemical potential vanishes and
$I_F[0]=0.3844$, the label $s$ is for sterile neutrinos produced via
a lepton-driven (MSW) resonance as described in ref.\cite{este}, and
the label $B$ stands for condensed or non-condensed Bosonic thermal
relics.

This expression features  several important consequences relevant
for large scale structure formation:
\begin{itemize}
 \item{A non-negligible $\nu_{wimp}\gg 10^{-12}$, for a CDM candidate
 with $m \sim 100\,\textrm{GeV}$ which decoupled at $T_d \sim 10 \,
 \mathrm{MeV}$\cite{dominik} overwhelms all other components (but for a possible BEC) and
 leads to   small free streaming lengths consistent with CDM  regardless of the presence of WDM or HDM components.
 For $\nu_{wimp} \gg 10^{-12}$ the free-streaming length of \emph{mixed} DM is completely dominated by WIMPs and
 is given today by \be \lambda_{fs}(0) \sim \frac{0.014\,\mathrm{pc}}{g^\frac{1}{3}_d}\,\Big( \frac{100\,\mathrm{GeV}}{m}\Big)^\frac{1}{2}\,
   \Big( \frac{10\,\mathrm{MeV}}{ T_d}\Big)^\frac{1}{2}\,, \ee from which it follows that for WIMPs with $m \sim
   100\,\mathrm{GeV}$ that decouple kinetically at $T_d \sim 10
   \,\mathrm{MeV}$\cite{dominik} when $g_d \sim 10$\cite{kt}  \be \lambda_{fs}(0) \sim   6.5\times
   10^{-3}\,\mathrm{pc}\,.\label{wimplum}\ee This cut-off scale might well be related
to the smallest non-linear structures found in\cite{cusps}   unless
there is some substantial violent relaxation and  merging. }

 \item{For vanishing chemical potential $u_d =0$, non-BEC Bosonic ultrarelativistic
 relics feature an infrared enhancement in  $I_B[x_d,0]$ which must be regulated by assuming that
 the integrals of the distribution function are cutoff of order of the Hubble scale at decoupling
 $H_d \sim \sqrt{g_d}\, T^2_d/M_{Pl}$.
  If this is the \emph{only} DM component, the
   free-streaming length  today is \be \lambda_{fs}(0) \sim
   \frac{14}{g^\frac{1}{3}_d\,\sqrt{I_B[x_d,0]}}\,\Big(
 \frac{\mathrm{keV}}{m}\Big) \mathrm{kpc}\,~~;~~I_B[x_d,0]  \sim
 \frac{1}{2\zeta(3)}\, \ln\Bigg[\frac{2\,T_d}{H_d+\sqrt{H^2_d+m^2}}\Bigg] \,   .\ee

    For BEC Bosons $x_d=u_d$ the
   function $I_B[x_d,x_d]$ is divergent as a consequence of the infrared divergence in the numerator of
   $I_a$ (eq. \ref{Ia}). However,  we highlighted that the physics of BEC formation
    in an expanding cosmology must be assessed in greater detail in order to understand the behavior of
    superhorizon modes. This observation also holds for the non-condensed Bose gas decoupling when relativistic
    with $\mu/T_d \ll1$ because the distribution function in this case is also infrared
    sensitive. In both these cases an estimate of the free streaming length may be obtained by introducing an infrared
    cutoff in the integral in the numerator of $I_a$
    of the order of the Hubble scale, since superhorizon modes cannot establish thermal equilibrium
    via causal processes as discussed above. If a BEC is the \emph{only} DM component,
    its free-streaming length today is approximately given by \be \lambda_{fs}(0) \sim
    \frac{0.014\,\mathrm{pc}}{g^\frac{1}{12}_d}\,\Big(\frac{\mathrm{keV}}{m} \Big)^\frac{3}{2}\,
    \Big(\frac{T_d}{\mathrm{GeV}} \Big)\,.  \ee Therefore even when these relics decoupled when they
    were ultrarelativistic, they could effectively act as CDM components. This is a consequence of the
    fact that the distribution functions favor small momenta. }

    \item{If sterile neutrinos that decouple \emph{out of LTE} near
    the QCD scale produced either non-resonantly via the (DW)\cite{dw} mechanism or
    via a lepton-driven MSW resonance\cite{este} near the QCD scale
    are the \emph{only} DM components we find the following
    free-streaming lengths today \bea   \lambda_{fs}(0)   & \simeq &  7  \, \mathrm{kpc}\,  \,\Big(
 \frac{\mathrm{keV}}{m}\Big)~~\mathrm{non-resonant}    \\\lambda_{fs}(0)  & \simeq &  1.73  \, \mathrm{kpc}\,  \,\Big(
 \frac{\mathrm{keV}}{m}\Big)~~\mathrm{resonant}\,.   \eea  The smaller values of the free streaming length
 are compatible with those in ref.\cite{gao} where it is found that first stars form  in filamentary
 structures with length scales of the order of the free streaming scale and within a factor $\sim 3-4$ seem to
 be also consistent with the ``cores'' resulting from the fit of the density profile for dwarf spheroidal
 galaxies in ref.\cite{gilmore} $\sim 0.5\,\mathrm{kpc}$. The larger values are consistent with those found
 in ref.\cite{salucci} for the cored profiles of spiral galaxies $\sim 5-10 \, \mathrm{kpc}$.    }

 \end{itemize}

 We believe that these results  lead to a significant advance in the understanding of
collisionless (DM) because trying to obtain a reliable estimate of
the free-streaming lengths   via the numerical integration of
Gilbert's equations for a combination of \emph{arbitrary}
distribution functions corresponding to particles that decoupled in
or out of LTE is undoubtedly a daunting task.

The sensitivity of the free-streaming scale to the details of the
distribution function at low momentum and the importance of a
reliable determination of the free-streaming length as a measure of
the cutoff of the power spectrum of linearized cosmological
perturbations require a fundamentally sound understanding of the
microphysics of production and decoupling of sterile neutrinos, a
program currently underway\cite{boyho}.

\acknowledgements{The author thanks H. J. de Vega and N. Sanchez for
stimulating discussions during the early stages of this work. He
acknowledges support
 from the U.S. National Science Foundation through grant No:  PHY-0553418.}

\end{document}